\documentclass{aastex631}

\usepackage{bm}
\usepackage{rotating}
\usepackage[hyphens]{url}
\usepackage{hyperref}

\hypersetup{linkcolor=red,citecolor=green,filecolor=cyan,urlcolor=magenta}
\shorttitle{Global analysis of the extended cosmic-ray decreases}
\shortauthors{Munakata et al.}
\graphicspath{{./}}

\begin{document}


\title{Global analysis of the extended cosmic-ray decreases observed with world-wide networks of neutron monitors and muon detectors; temporal variation of the rigidity spectrum and its implication}

\author[0000-0002-2131-4100]{K. Munakata}\affiliation{Physics Department, Shinshu University, Matsumoto, Japan}
\author[0000-0002-0890-0607]{Y. Hayashi}\affiliation{Physics Department, Shinshu University, Matsumoto, Japan}\author[0000-0002-3948-3666]{M. Kozai}\affiliation{Polar Environment Data Science Center, Joint Support-Center for Data Science Research, Research Organization of Information and Systems, Tachikawa, Japan}
\author[0000-0002-4913-8225]{C. Kato}\affiliation{Physics Department, Shinshu University, Matsumoto, Japan}
\author[0009-0006-3569-7380]{N. Miyashita}\affiliation{Physics Department, Shinshu University, Matsumoto, Japan}
\author[0000-0001-9400-1765]{R. Kataoka}\affiliation{National Institute of Polar Research, Tachikawa, Japan}
\author[0000-0002-6105-9562]{A. Kadokura}\affiliation{Polar Environment Data Science Center, Joint Support-Center for Data Science Research, Research Organization of Information and Systems, Tachikawa, Japan}\affiliation{National Institute of Polar Research, Tachikawa, Japan}
\author[0000-0002-3067-655X]{S. Miyake}\affiliation{National Institute of Technology (KOSEN), Gifu College, Gifu, Japan}
\author[0000-0002-2464-5212]{K. Iwai}\affiliation{Institute for Space-Earth Environmental Research, Nagoya University, Nagoya, Japan}
\author[0000-0002-8351-6779]{E. Echer}\affiliation{National Institute for Space Research, São José dos Campos, Brazil}
\author[0000-0002-4361-6492]{A. Dal Lago}\affiliation{National Institute for Space Research, São José dos Campos, Brazil}
\author[0000-0002-9737-9429]{M. Rockenbach}\affiliation{National Institute for Space Research, São José dos Campos, Brazil}
\author[0000-0002-7720-6491]{N. J. Schuch}\affiliation{National Institute for Space Research, São José dos Campos, Brazil}
\author[0000-0003-2931-8488]{J. V. Bageston}\affiliation{National Institute for Space Research, São José dos Campos, Brazil}
\author[0000-0003-1485-9564]{C. R. Braga}\affiliation{The Johns Hopkins University Applied Physics Laboratory, Laurel, MD, USA}
\author[0000-0002-4486-237X]{H. K. Al Jassar}\affiliation{Physics Department, Kuwait University, Kuwait City, Kuwait}
\author[0000-0002-2090-351X]{M. M. Sharma}\affiliation{Physics Department, Kuwait University, Kuwait City, Kuwait}
\author[0000-0001-7463-8267]{M. L. Duldig}\affiliation{School of Natural Sciences, University of Tasmania, Hobart, Australia}
\author[0000-0002-4698-1672]{J. E. Humble}\affiliation{School of Natural Sciences, University of Tasmania, Hobart, Australia}
\author[0000-0002-7037-322X]{I. Sabbah}\affiliation{Department of Applied Sciences, College of Technological Studies, Public Authority for Applied Education and Training, Shuwaikh, Kuwait}
\author[0000-0001-7929-810X]{P. Evenson}\affiliation{Bartol Research Institute, Department of Physics and Astronomy, University of Delaware, Newark, DE, USA}
\author[0000-0001-6584-9054]{T. Kuwabara}\affiliation{Bartol Research Institute, Department of Physics and Astronomy, University of Delaware, Newark, DE, USA}
\author[0000-0002-3715-0358]{J. K\'ota}\affiliation{Lunar and Planetary Laboratory, University of Arizona, Tucson, AZ 85721, USA}



\begin{abstract}
This paper presents the global analysis of two extended decreases of the galactic cosmic ray intensity observed by world-wide networks of ground-based detectors in 2012. This analysis is capable of separately deriving the cosmic ray density (or omnidirectional intensity) and anisotropy each as a function of time and rigidity. A simple diffusion model along the spiral field line between Earth and a cosmic-ray barrier indicates the long duration of these events resulting from about 190$^\circ$ eastern extension of a barrier such as an IP-shock followed by the sheath region and/or the corotating interaction region (CIR). It is suggested that the coronal mass ejection merging and compressing the preexisting CIR at its flank can produce such the extended barrier. The derived rigidity spectra of the density and anisotropy both vary in time during each event period. In particular we find that the temporal feature of the ``phantom Forbush decrease'' reported in an analyzed period is dependent on rigidity, looking quite different at different rigidities. From these rigidity spectra of the density and anisotropy, we derive the rigidity spectrum of the average parallel mean-free-path of pitch angle scattering along the spiral field line and infer the power spectrum of the magnetic fluctuation and its temporal variation. Possible physical cause of the strong rigidity dependence of the ``phantom Forbush decrease'' is also discussed. These results demonstrate the high-energy cosmic rays observed at Earth responding to remote space weather.
\end{abstract}

\keywords{Galactic cosmic rays (567) --- Forbush effect (546) --- Interplanetary shocks (829) --- Solar coronal mass ejections (1626) --- Interplanetary corotating streams (314)}

\section{Introduction} \label{sec:intro}

The galactic cosmic-ray (GCR) intensity observed at Earth dynamically changes following the arrival of interplanetary (IP) shocks produced by coronal mass ejections (CMEs) and/or the fast solar wind from coronal holes forming corotating interaction regions (CIRs). The GCRs, propagating through the heliosphere, are pushed back to the outer heliosphere by the magnetized solar wind plasma blowing away from the Sun resulting in a large-scale radial spatial gradient of the omnidirectional GCR intensity (GCR density). The IP-shock embedded in the solar wind passing across the Earth causes an additional temporal decrease of GCR density when Earth enters the GCR depleted region formed behind the IP-shock. The GCR density also decreases when Earth enters an expanding magnetic flux rope (MFR) in the CME behind the shock due to the adiabatic cooling inside the MFR. The Forbush decrease (FD) is the well-known example of such GCR decreases \citep{Wibberenz98}. In a typical FD, the GCR density decreases after the shock arrival and reaches at its minimum within a few days and then recovers to the original level within about a week. During the FD, the GCR directional anisotropy is also enhanced and observed as the enhanced diurnal variation of GCR intensity by a ground-based detector on the Earth. The enhanced diurnal anisotropy is mainly caused by the GCR diffusion along the interplanetary magnetic field (IMF) refilling the depleted region, in addition to the GCR streaming perpendicular to the IMF such as the diamagnetic flow and perpendicular diffusion.\par

The magnitude of the GCR density decrease in an FD is known to be inversely proportional to the mean-free-path of the pitch angle scattering which is responsible for the GCR diffusion \citep{Wibberenz98}. Since the mean-free-path increases with increasing GCR rigidity ($p$), the magnitude of the density decrease is expected to decrease with $p$ as often observed. The density decrease due to adiabatic cooling in an expanding MFR is also expected to be inversely proportional to $p$ (e.g. \citet{paper1}, hereafter referred as Paper 1). The amplitude of the diurnal anisotropy, on the other hand, is expected to only weakly depend on $p$, since the diffusion anisotropy is given by the spatial gradient of GCR density decreasing with $p$ multiplied by the scattering mean-free-path increasing with $p$. According to the quasi-linear-theory incorporating ``resonant scattering'' of cosmic rays gyrating around the IMF, the parallel mean-free-path is inversely proportional to the power of the magnetic fluctuation at the wave number which is inversely proportional to the particles' Larmor radii \citep{Jokipii66}. This implies that the rigidity spectrum of the scattering mean-free-path reflects the power spectrum of the magnetic fluctuation. It is expected, therefore, that one can infer the power spectrum of the IMF fluctuation by analyzing the rigidity dependencies of the GCR density and anisotropy. This paper aims at such analyses of two extended GCR decreases in 2012 described below.\par  

Recently, \citet{paper2} (hereafter referred to as Paper 2) analyzed two extended GCR decreases in July-August and January-February, 2012 shown in Figure \ref{fig:1hour}. About 5 \% decreases of the GCR intensity are observed by OULU neutron monitor (NM) and the duration of these events is very long  (see red curves in Figure \ref{fig:1hour}(d)). Following IP-shock arrivals identified by the onset of the geomagnetic storm sudden commencement (SSC)\footnote{http://isgi.unistra.fr} and indicated by orange vertical lines, the GCR density during these events takes a few weeks to reach its minimum and takes another few weeks to recover. Except those at shock arrivals, there is no notable enhancement in the IMF magnitude which seems to be capable of causing additional decreases in high-energy GCR intensity and/or impeding recoveries from those decreases. Another notable feature of these events is the strong enhancement of the GCR anisotropy which is seen as the large amplitude diurnal variations in the count rate. An interesting feature of the event in January-February, 2012 is the OULU NM data (the red curve in the right panel of Figure \ref{fig:1hour}(d)) showing the second decrease starting at the day of year (DOY) 31 when no significant enhancement in the solar wind parameters is observed. \citet{Thomas15} referred to this ``phantom FD'', as evidence for GCR modulation by remote solar wind structures which are not observed at Earth. It is important to note in this figure that the second decrease is not seen in the PSNM NM (black curve) and the Nagoya-MD (blue curve), both monitoring higher rigidity GCRs than OULU NM. This suggests that the temporal variation of the GCR intensity in this event is strongly dependent on the GCR rigidity.\par

By analyzing the diurnal variations of GCR intensity observed by two NMs at low and high geomagnetic cutoff rigidities, Paper 2 reported that the strong anisotropy is consistent with the so-called ``corotation anisotropy'' from 18:00 local time (LT) which is expected from the parallel diffusion from 21:00 LT along the nominal Parker's IMF added to the radial solar wind convection anisotropy from 12:00 LT. They attributed the long duration of these event to a large extent of the IP-shock barrier responsible to the GCR decrease far to east of the Sun-Earth line (viewed from Earth), due to this Earth remains magnetically connected to the shock for a long period. They also concluded that these two events are weakly dependent on GCR rigidity and suggested magnetic mirroring as a possible mechanism responsible for such rigidity independent barrier.\par

However, it is difficult to accurately analyze an FD and its rigidity dependence by analyzing only a few NMs data, particularly when the strong anisotropy is superposed on the observed temporal variation of the GCR density. Since the GCR variation observed at a point in space generally consists of the variation of the GCR density and the variation due to the anisotropy, the multidirectional observations using a global detector network are necessary to study these components separately and accurately. For this purpose, the world-wide network observations with the ground-based detectors have been employed on the global analyses of the GCR intensity variation. Examples of such analyses using the neutron monitor network data can be found in \citet{Belov18} and \citet{Abunin20}. The neutron monitors, which detect secondary neutrons produced by GCRs interacting with atmospheric nuclei, have a maximum response to primary GCRs with the median rigidities between $\sim10$ GV and $\sim30$ GV. Muon detectors, on the other hand, detect muons produced through the hadronic interaction between GCRs and atmospheric nuclei. Because a higher primary energy is needed to produce muons with sufficient Lorentz factor and relativistic time dilation to reach ground level before decaying, muon detectors (MDs) have a response to primary GCRs with higher median rigidities, between $\sim50$ GV and $\sim100$ GV. While the NM is an omnidirectional detector, its observations are dominated by the vertically incident GCRs to the detector, a single MD can be multidirectional because the incident direction of muons better preserves the incident direction of primary GCRs at the top of the atmosphere. The Global Muon Detector Network (GMDN) was established in 2006 with four multidirectional surface muon detectors at Nagoya in Japan, Hobart in Australia, Kuwait City in Kuwait and S\~ao Martinho in Brazil \citep{Okazaki08}. By analyzing data observed with NMs and MDs each forming a global network, therefore, it becomes possible to accurately deduce the rigidity dependence of an FD over an extended rigidity range.\par

In this paper, we apply this global analysis method to two extended decreases shown in Figure \ref{fig:1hour}. Paper 1 has performed a similar analysis, though only for a three day period when a large-amplitude bidirectional anisotropy is observed in an FD in November, 2021, and demonstrated the rigidity dependencies of the GCR density and anisotropy dynamically changing during the FD. For an analysis of the GCR decrease extended over longer period such as one solar rotation period, one needs to suppress the influences of ``local effects''. In this paper, we apply the global analysis method improved in this way, as described in Appendix \ref{sec:appendix}.\par

We describe the data and analysis in Sections \ref{subsec:data} and \ref{subsec:analyses}, respectively, and show the results in Section \ref{sec:result}. Finally, the discussion and summary are given in Section \ref{sec:discussion}.\par

\section{Data and analyses} \label{sec:DandA}
To derive the rigidity spectra of the cosmic-ray density and anisotropy, we analyze hourly count rates recorded by worldwide networks of NMs and MDs which record GCR intensities in various viewing directions and in different rigidity regions. In the following subsections we describe the analyzed data and the analysis method.\par

\subsection{Analyzed periods and data} \label{subsec:data}
We analyze the extended cosmic-ray decreases recorded during two solar rotation periods (27-day periods) from 13 July (DOY 195) to 9 August (DOY 221) and from 17 January (DOY 17) to 13 February (DOY 43), 2012, which are hereafter referred as Period I and Period II, respectively. Figure \ref{fig:1hour} displays hourly count rates of sample NMs and MDs together with solar wind parameters downloaded from the omni-website\footnote{https://omniweb.gsfc.nasa.gov/ow.html \label{footnote_2}}.\par

In this paper, we use hourly count rates recorded by 20 NMs and 60 directional channels of GMDN (hereafter 60 MDs). Characteristics of these NMs and MDs are given in Table 1 of Paper 1. Because the observations with an NM and a MD at the Syowa station started in 2018, those data used in Paper 1 are not available for the analysis of FD events in 2012. On the other hand, we include data from six additional NMs in the analysis of this paper. Table 1 lists characteristics of these additional NMs used in this paper. Thus, here we use 86 hourly count rates by 26 NMs and 60 directional channels of GMDN altogether, which are available at websites\footnote{https://www01.nmdb.eu/} \footnote{http://hdl.handle.net/10091/0002001448}. All cosmic-ray data are corrected for the atmospheric pressure effect, while we added an additional correction to GMDN data for the atmospheric temperature effect by applying the method developed by \citet{Rafael16}. This method uses the mass weighted temperature calculated from the vertical profile of the atmospheric temperature provided by the Global Data Assimilation System (GDAS) of the National Center for Environmental Prediction available at the NOAA website\footnote{https://www.ready.noaa.gov/gdas1.php}.\par

\subsection{Global Analysis of Cosmic-Ray data} \label{subsec:analyses}
We analyze the normalized percent deviation $I^{nor}_{i,j}(t)$ calculated from the hourly count rate of the $j$-th directional channel of the $i$-th detector ($j=1$ for all NMs, $j=1-17$ for Nagoya and São Martinho da Serra MDs, $j=1-13$ for Hobart and Kuwait MDs) at universal time $t$ in hour. Our analysis is identical to that described in Paper 1 and readers are referred to that paper for the details of the procedure. In the analysis of this paper, the influence of the local effects are suppressed in $I^{nor}_{i,j}(t)$ (see Appendix \ref{sec:appendix}).\par

Cosmic ray decreases in Period I and II are accompanied by strong geomagnetic storms which slightly change the geomagnetic cut-off rigidity ($P_c$) and the asymptotic direction ($\lambda_{asymp}, \phi_{asymp}$) in Table 1. However, the influence of these changes on the overview of the best-fit parameters discussed in this paper is small \citep{Munakata18}.\par

The hourly best-fit parameters derived in Paper 1 showed dynamic temporal variations even in only a 3-day period, responding to varying solar wind parameters. In the present analysis of cosmic-ray decreases extending over much longer periods, therefore, we review more gradual variations of comic-ray density ($\xi_c^{0,0}(t)$), the diurnal anisotropy vector $\bm{\xi}_1(t)=(\xi_c^{1,0}(t), \xi_c^{1,1}(t), \xi_s^{1,1}(t))$ and their rigidity spectra, using the 24-hour central moving average (CMA) calculated by Eq.\ref{24hCMA} in Appendix \ref{sec:appendix} from hourly values. In the next section, we present 24-hour CMAs of these cosmic-ray parameters in comparison with the solar wind parameters.\par

\section{Results}\label{sec:result}
Figures \ref{fig:bfparam}(a) and \ref{fig:bfparam}(b) show the 24-hour CMA of hourly solar wind parameters observed during Period I (left) and Period II (right), while Figures \ref{fig:bfparam}(c)-\ref{fig:bfparam}(f) display the 24-hour CMA of hourly best-fit parameters obtained from the global analysis described in the preceding section, each with 6 hour cadence. In Figure \ref{fig:bfparam}(a), significant enhancements are seen in the IMF magnitude (red curve on the left vertical axis) and the solar wind velocity (blue curve on the right vertical axis) at the IP-shock arrival indicated by vertical orange lines. Figure \ref{fig:bfparam}(b) shows the GSE-longitude (red curve) and latitude (blue curve) of the IMF orientation which are calculated from 24-hour CMAs of three IMF components, each with an error deduced from the dispersion of hourly IMF-longitudes and latitudes in 24-hours used for calculating CMAs. The GSE-longitude in each period is overall consistent with the nominal Parker's spiral field in the ecliptic plane with multiple IMF-sector boundaries. Figure \ref{fig:bfparam}(c) shows the cosmic-ray density in space ($\bar{\xi}_c^{0,0}(t_i)$) at 15 GV (red curves) and 65 GV (blue curves). We calculate the best-fit parameters at 15 GV and 65 GV by setting the reference rigidity $p_r$ in Eq.3 of Paper 1 to 15 GV and 65 GV, respectively. In order to display on a common vertical axis, $\bar{\xi}_c^{0,0}(t_i)$ at 65 GV in this panel are multiplied by 65/15 by assuming $1/p$ rigidity dependence, while gray curves show the residual after subtracting blue curves from red curves on the right vertical axes indicating the relative excess density at 15 GV. Since the daily mean count rate of each detector is normalized to 0~\% at the first day of each period as described in Appendix \ref{sec:appendix}, the zonal components $\bar{\xi}_c^{n,0}(t_i)$ including the density $\bar{\xi}_c^{0,0}(t_i)$ in this panel represent the variation relative to the first day. It is clearly seen that the cosmic-ray decrease starts at the IP-shock arrival and lasts for a few weeks. As shown by black curves in Figure \ref{fig:bfparam}(d), the amplitude of the diurnal anisotropy ($\vert\bm{\xi}(t)\vert$) is enhanced in both periods with maxima far larger than the so-called ``corotation anisotropy'' in quiet periods with an average amplitude of $\sim$ 0.5~\% \citep{Munakata14}. In the following subsections, we briefly review best-fit parameters in each period.\par

\subsection{Cosmic ray decrease in July-August, 2012}\label{subsec:july2012}
In Period I, $\bar{\xi}_c^{0,0}(t)$ at 15 GV shown by the red curve in Figure \ref{fig:bfparam}(c) decreases during the first three days (DOY 197-199) and recovers afterward. This first decrease is caused by an expanding magnetic flux rope (MFR) indicated by a smooth rotation of a strong IMF in DOY 197-198 as seen in Figures \ref{fig:1hour}(a) and \ref{fig:1hour}(b). An interplanetary counterpart of the CME (ICME) has arrived at Earth on 18:09 UT of DOY 196 (July 14) and caused the SSC\footnote{https://dataverse.harvard.edu/dataset.xhtml?persistentId=doi:10.7910/DVN/C2MHTH}\citep{Rich24}. $\bar{\xi}_c^{0,0}(t)$ decreases again toward its minimum at around DOY 204 and then recovers throughout the rest of the plotted period. There is another dip seen in $\bar{\xi}_c^{0,0}(t)$ at around DOY 210 in the recovery phase when an IMF enhancement is observed. IMF-sector boundaries are also observed during this dip period (see red curve in Figure \ref{fig:bfparam}(b)). As shown by blue curve in Figure \ref{fig:bfparam}(c), temporal variations of $\bar{\xi}_c^{0,0}(t)$ at 65 GV generally looks similar in this event. This is also indicated by $\bar{\gamma}_0(t)$ shown by red curve in Figure \ref{fig:bfparam}(f) which is almost constant at around -1.0 for 10 days between DOY 199 and DOY 210. The difference between red and blue curves shown by gray curve in \ref{fig:bfparam}(c) also varies gently between -1 \% and 0 \%. On the other hand, $\bar{\gamma}_1(t)$ shown by blue curve varies between -0.5 and 0.0 in the same period indicating that the spectrum of the diurnal anisotropy is systematically harder than that of the cosmic-ray density.\par

As shown by the black curve in Figure \ref{fig:bfparam}(d), the amplitude of the diurnal anisotropy ($\vert\bar{\bm{\xi}}(t)\vert=\sqrt{{\bar{\xi}_{\parallel}(t)}^2+{\bar{\xi}_{\perp}(t)}^2}$) in this event reaches about 2~\% at 15 GV. It is interesting to note that the anisotropy amplitude displays two minima at around DOY 198 and DOY 210 when the cosmic-ray density displays local minima in Figure \ref{fig:bfparam}(c). Red and blue curves in this figure show magnitudes of the anisotropy components parallel ($\bar{\xi}_{\parallel}(t)$) and perpendicular ($\bar{\xi}_{\perp}(t)$) to the IMF, respectively. $\bar{\xi}_{\parallel}(t)$ tends to dominate over $\bar{\xi}_{\perp}(t)$ in this event. The GSE-longitude of the anisotropy orientation shown by red curve in Figure \ref{fig:bfparam}(e) indicates that the anisotropy orientation is in the second quadrant of the GSE-coordinate as expected from the cosmic-ray streaming from the outer heliosphere beyond Earth along the IMF field line. Interestingly, the GSE-latitude of the anisotropy orientation (blue curve) is positive during the most of the period indicating cosmic-rays predominantly streaming from the north of Earth.\par

\subsection{Cosmic ray decrease in January-February, 2012}\label{subsec:january2012}
In Period II, two ICMEs are reported on DOY 21 and DOY 24 responsible to the first and second SSCs\footnote{https://helioforecast.space/icmecat \label{footnote_7}}. It is seen in Figure \ref{fig:bfparam} that temporal variations of $\bar{\xi}_c^{0,0}(t)$ at 15 GV and 65 GV look different in DOY 26-37. As shown in Figure \ref{fig:bfparam}(c), they decrease similarly until DOY 26, but $\bar{\xi}_c^{0,0}(t)$ at 15 GV (red curve) then stops decreasing until DOY 30, while $\bar{\xi}_c^{0,0}(t)$ at 65 GV (blue curve) continues decreasing. After DOY 31, $\bar{\xi}_c^{0,0}(t)$ at 15 GV resumes decreasing and then recovers earlier than $\bar{\xi}_c^{0,0}(t)$ at 65 GV. Thus, $\bar{\xi}_c^{0,0}(t)$ at 15 GV looks like decreasing in two steps with the second decrease during DOY 31-36. As seen Figure \ref{fig:1hour}(d), this second decrease is less evident in the 24-hour CMAs observed by PSNM and MDs which are monitoring higher rigidity cosmic rays. As mentioned in Section \ref{sec:intro}, \citet{Thomas15} analyzed this event observed by four high-latitude NMs and concluded that the second decrease at DOY 31 in this event lacks obviously modulating structure in near-Earth space. They referred to this event as a ``phantom FD''.\par


The spectral index ($\bar{\gamma}_0(t)$) shown by the red curve in Figure \ref{fig:bfparam}(f) represents the rigidity dependence of $\bar{\xi}_c^{0,0}(t)$. Following large variation before DOY 26 when $\bar{\xi}_c^{0,0}(t)$ at 15 GV and 65 GV decrease similarly, $\bar{\gamma}_0(t)$ deviates from -1.0 up to -0.5 displaying two maxima at $\sim$ DOY 29 and $\sim$ DOY 37 when the difference between red and blue curves becomes most prominent in Figure \ref{fig:bfparam}(c). It is noted that $\bar{\gamma}_0(t)$ is not the power-law index of the GCR intensity spectrum which increases according to less depression in the higher rigidities. Instead, $\bar{\gamma}_0(t)$ is the spectral index of the magnitude of the relative intensity depression which increases according to more depression in the higher rigidity intensity (see Paper 1).\par

The maximum amplitude of the diurnal anisotropy ($\vert\bar{\bm{\xi}}(t)\vert$) in this event also exceeds 1.5 \% as shown by black curve in Figure \ref{fig:bfparam}(d) with similar contributions from $\bar{\xi}_{\parallel}(t)$ and $\bar{\xi}_{\perp}(t)$. The GSE-longitude of the anisotropy orientation is roughly consistent with the cosmic-ray streaming from the outer heliosphere along the IMF. $\vert\bar{\bm{\xi}}(t)\vert$ decreases at around DOY 33 near the cosmic-ray density minimum in Figure \ref{fig:bfparam}(c). Interestingly, an IMF-sector boundary is observed during this minimum at DOY 32-33 as seen in Figure \ref{fig:bfparam}(b). $\bar{\gamma}_1(t)$ also varies between -0.5 and 0.0 in this event.\par

In the next section, we discuss the observed features of cosmic-ray density and anisotropy shown above, based on a simple two-dimensional diffusion model.\par

\section{Discussion and conclusions} \label{sec:discussion}
The most unusual feature of two cosmic-ray decreases analyzed in the preceding section is their extremely long duration reaching 27 days. A single strong IP-shock accompanied by a CME and/or MFR can cause an FD, but the cosmic-ray decrease and its recovery usually ends within a week or so. The long duration of two events might be caused by multiple IP-shocks successively hitting Earth during a longer period, but there is no clear signature of such successive shocks and/or ICMEs capable of keeping the high energy GCR intensity depleted for so long.\par 

\begin{figure}[ht!]
\plotone{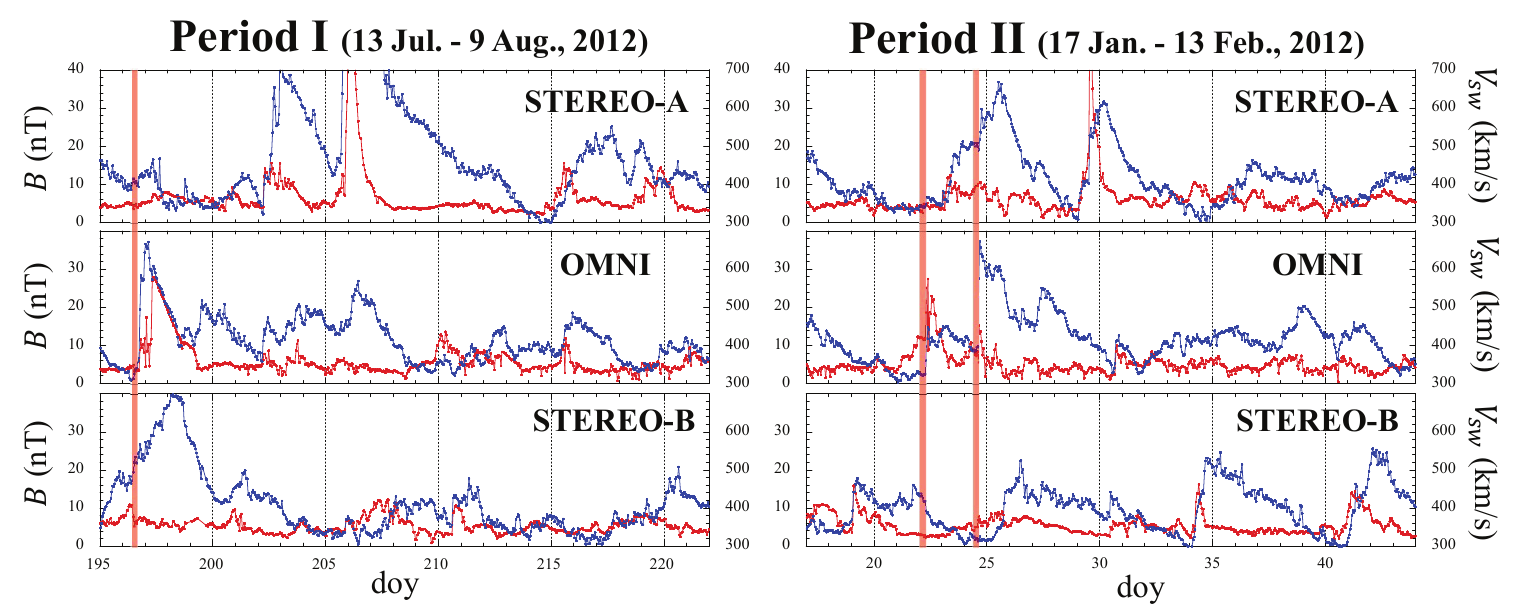}
\caption{Hourly IMF magnitude and solar wind velocity measured by \textit{STEREO-A} and \textit{STEREO-B} and those measured at Earth during Period I (left) and Period II (right). In each panel, red curve shows the IMF magnitude on the left vertical axis, while blue curve shows the solar wind velocity on the right vertical axis. The \textit{STEREO-A} and \textit{STEREO-B} are located at about 120$^\circ$ west and east of the Sun-Earth line (viewed from Earth) in both periods, respectively. The times of the IP-shock arrival at Earth identified by the SSC onset are indicated by vertical orange lines.\label{fig:stereo}}
\end{figure}

\begin{figure}[ht!]
\epsscale{.6}
\plotone{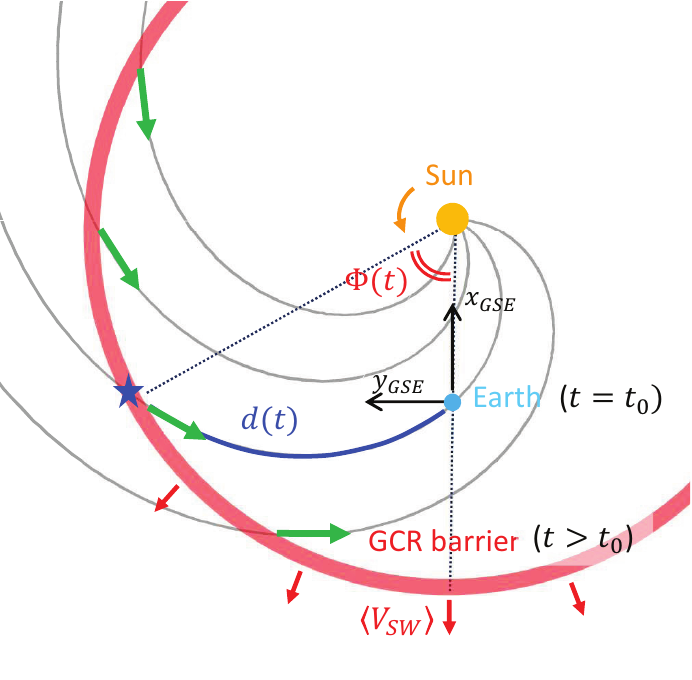}
\caption{Illustration of Earth connected with the GCR barrier along the IMF field line. This figure illustrates the spherical GCR barrier such as IP-shock followed by the sheath region and/or the CIR moving radially outward from Sun at the time $t$, when Earth is connected with the intersection on the barrier indicated by a blue star along an IMF field line depicted by a blue curve. The barrier arrived at Earth at $t_0<t$. As the barrier propagates away from Sun, the intersection moves eastward (viewed from Earth) on the barrier at the distance $d(t)$ measured along the field line and at an angle $\Phi(t)$ measured from the Sun-Earth line. Green arrows indicate the cosmic-ray streaming which is observed at Earth as a parallel anisotropy along the field line. This figure is not drawn to scale.\label{fig:model}}
\end{figure}

The extended GCR decreases in Periods I and II are closely related to CMEs at this time. Figure \ref{fig:stereo} shows the hourly IMF magnitude and solar wind velocity observed by \textit{STEREO-A} (top panels) and \textit{STEREO-B} (bottom panels) in Period I and Period II$^{\ref{footnote_2}}$. During these periods, \textit{STEREO-A} and \textit{STEREO-B} are located at about 120$^\circ$ west and east of the Sun-Earth line, respectively. According to the Space Weather Database Of Notification, Knowledge, Information (DONKI)\footnote{https://kauai.ccmc.gsfc.nasa.gov/DONKI/search/ \label{footnote_8}}, the CME responsible for the SSC near the beginning of Period I occurred on DOY 194 (July 12) and propagated along the solar equatorial plane with a large longitudinal extension, as seen in the plasma density movie in the Enlil time dependent magnetohydrodynamics model\footnote{\url{https://iswa.gsfc.nasa.gov/downloads/20120712_193500_anim.tim-den.gif}}. After passing Earth, this CME merged and compressed a preexisting CIRs at its flank. This merging formed a long interaction region extended over eastern longitudes. \citet{Thomas15} also noted such CME merging with the preexisting CIR. These CIRs are seen as velocity enhancements at \textit{STEREO-B} and \textit{STEREO-A} on DOY 197-198 (July 15-16) and on DOY 202-203 (July 20-21), respectively, in Figure \ref{fig:stereo}. It is known that the CIR modulating GCRs develops beyond Earth's orbit with steepened forward and reverse shocks \citep{Rich04}. It is likely therefore that these long-lasting compressed CIRs are responsible for the extended GCR decrease in Period I, working as barriers for GCR propagation to Earth along the IMF. There is another fast CME observed by \textit{STEREO-A} on DOY 205 (July 23) as seen in Figure \ref{fig:stereo}, but it did not arrive at Earth and \textit{STEREO-B}\footnote{\url{https://iswa.gsfc.nasa.gov/downloads/20120723_033000_anim.tim-den.gif}}\citep{Liu14}. In Period II, there were two CMEs erupting on DOY 19 and DOY 23$^{\ref{footnote_8}}$. These CMEs again merged and compressed the preexisting CIRs at their eastern flanks forming the extended interaction regions over the eastern longitude of Earth\footnote{\url{https://iswa.gsfc.nasa.gov/downloads/20120123_052000_anim.tim-den.gif}}. The preexisting CIRs can be seen as the velocity enhancements at \textit{STEREO-A} and \textit{STEREO-B} in Figure \ref{fig:stereo}. Thus, CMEs in Period I and II interact with the preexisting CIRs at its flank forming the enhanced interaction regions which possibly operate as the barrier for GCR propagation to Earth (the GCR barrier) for a long time and cause the extended GCR decrease observed at Earth. Now we discuss how the observed results can be expected from the GCR barrier based on a simple two-dimensional diffusion model illustrated in Figure \ref{fig:model}.\par

\begin{figure}[ht!]
\plotone{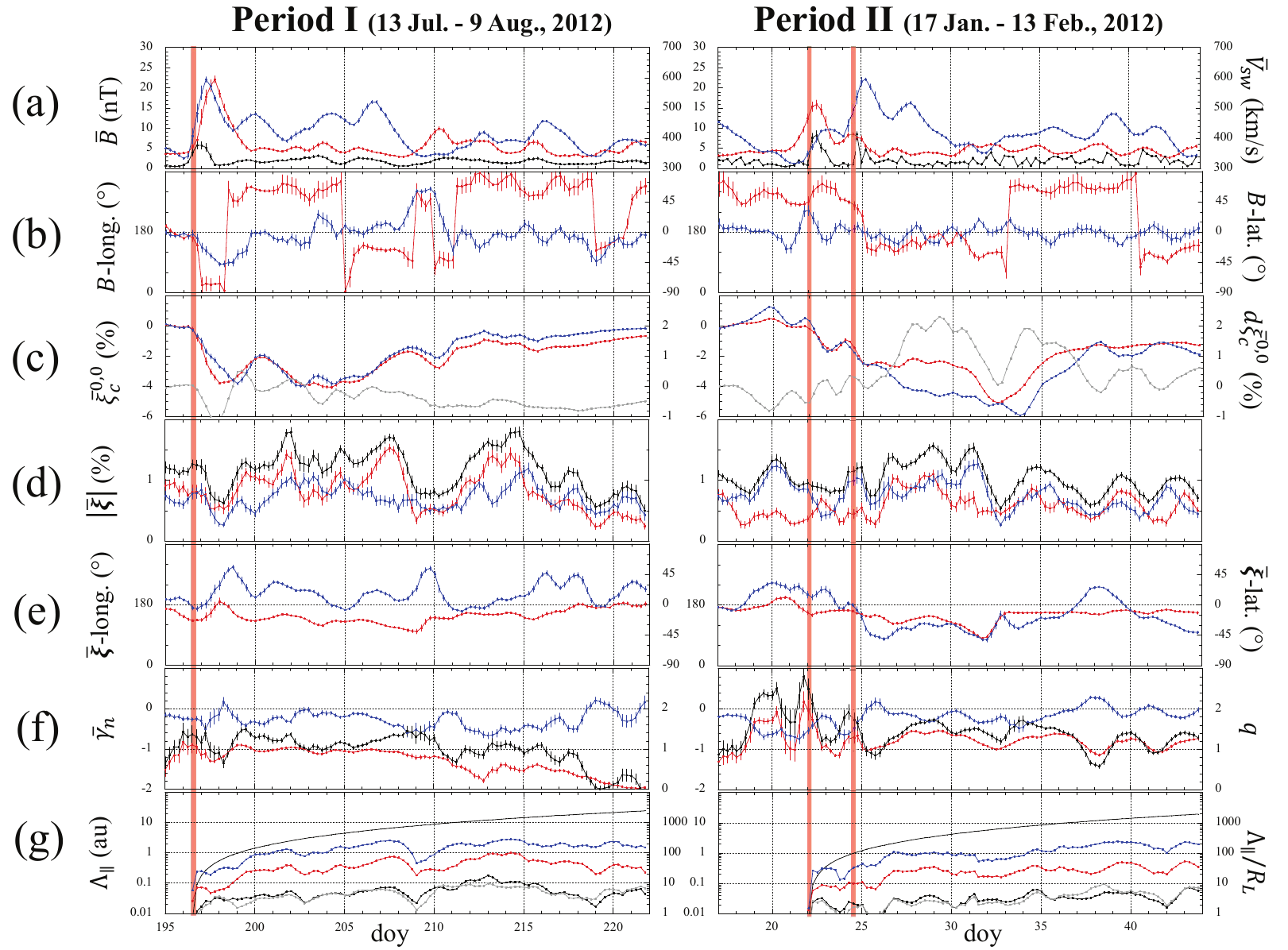}
\caption{The 24-hour central moving averages (CMAs) of solar wind parameters and cosmic ray parameters. In each panel, the 24-hour CMA calculated by Eq.\ref{24hCMA} is plotted with 6 hour cadence with the error deduced from the dispersion of 24 hourly data used to calculate CMA. Panels (a) and (b) show the solar wind parameters in the same format as Figures 1(a) and 1(b). Panels (c)-(g) show cosmic ray parameters; (c) best fit GCR density $\bar{\xi}_c^{0,0}(t)$ at 15 GV (red curves) and 65 GV (blue curves) on the left and right vertical axes, respectively, (d) amplitudes of the best-fit diurnal anisotropy vector $\vert\bar{\bm{\xi}}(t)\vert$ (black curves), parallel anisotropy $\vert\bar{\xi}_{\parallel}(t)\vert$ (red curves) and perpendicular anisotropy $\vert\bar{\xi}_{\perp}(t)\vert$ (blue curves), all at 15 GV, (e) GSE-longitude (red curves) and latitude (blue curves) of $\bar{\bm{\xi}}(t)$'s orientation on the left and right vertical axes, respectively. In panel (f), power-law indices of the best-fit rigidity spectra, $\bar{\gamma}_0(t)$ (red curves) and $\bar{\gamma}_1(t)$ (blue curves), are plotted on the left vertical axes, while the inferred power-law index $\bar{q}(t)=2-(\bar{\gamma}_1(t)-\bar{\gamma}_0(t))$ of the magnetic field fluctuation are plotted by black curves on the right vertical axes. In panel (g), the distance $d(t)$ along the field line between Earth and the intersection on the GCR barrier (thin black curve), the average parallel mean-free-paths of pitch-angle scattering $\Lambda_\parallel(t)$ at 15 GV (red curves) and 65 GV (blue curves) are plotted on the left vertical axes, while $\Lambda_\parallel(t)/R_L(t)$ at 15 GV (black curves) and 65 GV (gray curves) are plotted on the right vertical axes. In panel (c), $\bar{\xi}_c^{0,0}(t_i)$ at 65 GV is multiplied by 65/15 by assuming $1/p$ rigidity dependence in order to display on a common vertical axis. Gray curves in this panel show the residual after subtracting blue curves from red curves on the right vertical axes indicating the excess density at 15 GV relative to that at 65 GV (see text). The times of the IP-shock arrival identified by the SSC onset are indicated by vertical orange lines.\label{fig:bfparam}}
\end{figure}

This figure depicts a partial spherical GCR barrier with a finite angular extension such as an IP-shock followed by the sheath region and/or CIR moving radially away from Sun at the time $t$ with an average solar wind velocity $\langle V_{sw} \rangle$ in the ecliptic plane and forming the cosmic-ray depleted region downstream of the barrier. When Earth enters this region, cosmic-ray detectors on Earth observe an FD. Cosmic rays upstream the barrier diffuse into the depleted region along a field line (gray curves) through the pitch-angle scattering by the irregular components of the magnetic field. The cosmic-ray streaming (green arrows) is observed at Earth as a parallel anisotropy along the field line connecting Earth with an intersection on the barrier marked by a blue star symbol. As the barrier propagates away from Earth, the intersection moves eastward (viewed from Earth) to a different point on the barrier. The distance $d(t)$ along the spiral field line between Earth and the intersection is given as
\begin{equation}
\label{d}
d(t)=\int_{r_E}^{r_E+\langle V_{sw} \rangle (t-t_0)} \sqrt{1+(r\Omega/ \langle V_{sw} \rangle)^2}dr
\end{equation}
where $r_E$ is 1 au, $r$ is the radial distance from Sun, $\Omega$ is the angular velocity of Sun viewed from Earth and $t-t_0$ is the time elapsed after Earth's crossing the barrier at $t_0<t$. Thin black curves in Figure \ref{fig:bfparam}(g) (bottom panels) show $d(t)$ calculated by using $\langle V_{sw} \rangle$, which is 419.9 km/s and 418.8 km/s in Period I and Period II, respectively. When $d(t)$ reaches 10 au, the radial distance of the barrier from Sun is about 4.5 au and the eastward angular deviation of the intersection from the Sun-Earth line (see $\Phi$ in Figure \ref{fig:model}) is calculated to be about 190\degr. This gives a rough estimate of the eastern angular extension of barriers responsible for the extended GCR decreases in Periods I and II. The eastern extensions of the compressed CIRs in Enlil movies seems to be not contradicting this estimate.\par

As shown in Figure \ref{fig:bfparam}(d), anisotropy components $\bar{\xi}_{\parallel}(t)$ and $\bar{\xi}_{\perp}(t)$ are both enhanced. We first discuss $\bar{\xi}_{\parallel}(t)$ based on the model in Figure \ref{fig:model}. The average magnitude of $\bar{\xi}_{\parallel}(t)$ is proportional to the average density gradient along the field line between Earth and the intersection and can be approximated as
\begin{equation}
\label{xipara}
\bar{\xi}_{\parallel}(t)=\Lambda_\parallel(t)\{\Delta{\bar{\xi}_c^{0,0}(t)}/d(t)\}
\end{equation}
where $\Delta{\bar{\xi}_c^{0,0}(t)}$ is the difference between $\bar{\xi}_c^{0,0}(t)$ at Earth and the intersection at a distance $d(t)$ from Earth and $\Lambda_\parallel(t)$ and $\Delta{\bar{\xi}_c^{0,0}(t)}/d(t)$ are the average parallel mean-free-path and density gradient along the field line, respectively. By using the observed $\bar{\xi}_{\parallel}(t)$ in Eq.\ref{xipara}, we evaluate $\Lambda_\parallel(t)$ as
\begin{equation}
\label{lpara}
\Lambda_\parallel(t)=\bar{\xi}_{\parallel}(t)d(t)/\Delta{\bar{\xi}_c^{0,0}(t)} 
\end{equation}
where $\Delta{\bar{\xi}_c^{0,0}(t)}$ is calculated from the observed $\bar{\xi}_c^{0,0}(t)$ as
\begin{equation}
\label{Dxi}
\Delta{\bar{\xi}_c^{0,0}(t)}=\{\bar{\xi}_c^{0,0}(t_0)+\langle G_r \rangle \langle V_{sw} \rangle (t-t_0) \}-\bar{\xi}_c^{0,0}(t), 
\end{equation}
by assuming that the cosmic-ray density is unmodulated at the intersection. $\langle G_r \rangle$ in Eq.\ref{Dxi} is the average large-scale radial density gradient which is calculated from the observed anisotropy later in this section (in period I, it is 4.32 \%/au and 0.67 \%/au at 15 GV and 65 GV, respectively, while it is 4.96 \%/au and 0.78 \%/au in Period II).  We set $t_0$ at DOY 196.5 and DOY 22.0 in Periods I and II, respectively, as indicated by the leftmost vertical orange lines in Figure \ref{fig:bfparam}.\par

Red and blue curves in Figure \ref{fig:bfparam}(g) display $\Lambda_\parallel(t)$ calculated at 15 GV and 65 GV, respectively. The ratios of $\Lambda_\parallel(t)$ to the particles Larmor radius ($R_L(t)=p/(cB(t)))$ at 15 GV and 65 GV are also shown by black and gray curves on the right vertical axis, respectively. It is seen that $\Lambda_\parallel(t)$ first rapidly increases as the GCR barrier moves away from Earth and then gradually increases exhibiting local minima and maxima as $d(t)$ increases. Although $\Lambda_\parallel(t)$ is proportional to $d(t)$ in Eq.\ref{lpara}, its increase with $d(t)$ is not trivial, because it is calculated by using the observed $\bar{\xi}_{\parallel}(t)$ and $\Delta{\bar{\xi}_c^{0,0}(t)}$, both changing in time.\par


$\Lambda_\parallel(t)$ is given in Eq.\ref{lpara} by $\bar{\xi}_{\parallel}(t)$ proportional to $p^{\bar{\gamma}_1(t)}$ divided by $\Delta{\bar{\xi}_c^{0,0}(t)}$ which is roughly proportional to $p^{\bar{\gamma}_0(t)}$. The power-law spectral index of $\Lambda_\parallel(t)$ is therefore given by $\bar{\gamma}_1(t)-\bar{\gamma}_0(t)$. In period II, this index decreases from about 1.0 to 0.3 in DOY 26-37 suppressing the increase of $\Lambda_\parallel(t)$ with $p$. Thus, it is indicated by the two-dimensional diffusion model that this ``softening'' of the rigidity dependence of $\Lambda_\parallel(t)$ is responsible for the observed ``phantom FD'' in Period II. Since the radial distance of the GCR barrier in DOY 26-37 is between 1.4 au and 4.3 au from Sun and $d(t)$ and $\Phi(t)$ in Figure \ref{fig:model} are between 2.4 au and 13.4 au and between 66$^\circ$ and 79$^\circ$, respectively, the model indicates that the ``softening'' of $\Lambda_\parallel(t)$ occurred when Earth is connected by the field line with this region on the barrier, although the physical mechanism responsible for the ``softening'' is unknown.\par

In Period II, multiple CMEs erupted from the Sun with various speeds including two which arrived at Earth on DOY 21 and DOY 24 as mentioned in Section\ref{subsec:january2012}$^{\ref{footnote_8}}$. It is possible, therefore, to expect that the faster CME overtook the preceding slower CME and compressed the magnetic field in between at the radial distance between 1.4 au and 4.3 au. Such compression can accelerate GCRs by the adiabatic heating causing the increase of GCR density which is inversely proportional to $p$. The increase of 1 \% needs only $\Delta E/E = 0.2-0.3$ \% range, that is few tens of MeVs  for 15 GeV GCRs. This density increase shows as an additional unmodulated cosmic-ray density (the first term on the right hand side of Eq.\ref{Dxi}) particularly for low-energy GCRs. The gray curve in the right panel of Figure \ref{fig:bfparam}(c) for Period II displays positive residual in DOY 26-37 indicating the excess GCR density at 15 GV relative to that at 65 GV. The gray curve shows two bumps separated by the IMF sector boundary observed in DOY 32-33 in Figure \ref{fig:bfparam}(b), probably indicating the adiabatic heating in two regions separated by the sector boundary. Based on the observation with the GMDN, \citet{Kihara21} reported a short bump of GCR density which is consistent with the adiabatic heating near the trailing edge of the flux rope observed in August 2018. If this is the case, the ``softening'' of $\Lambda_\parallel(t)$ mentioned above is not necessary for interpreting the ``phantom FD'' in Period II. The interaction between CMEs are also expected in Period I, but its influence on GCR desnity is less significant as seen in the gray curve in the left panel of Figure \ref{fig:bfparam}(c) showing only a gentle variation.\par

According to the quasi-linear-theory of the diffusive GCR propagation incorporating ``resonant scattering'' of cosmic rays in an irregular magnetic field, the parallel mean-free-path for cosmic rays gyrating at a pitch-angle cosine $\mu$ with the rigidity $p$ is inversely proportional to the power of the magnetic fluctuation at the wave number $k=1/(R_L\vert\mu\vert)$ \citep{Jokipii66}. If the power spectrum of the magnetic fluctuation is expressed by $k^{-q(t)}$, $\Lambda_\parallel(t)$ is expected to be proportional to $p^{2-q(t)}$ and $q(t)$ is given by $q(t)=2-(\bar{\gamma}_1(t)-\bar{\gamma}_0(t))$. In Figure \ref{fig:bfparam}(f), $q(t)$ is displayed by the black curve with the right vertical axis. It varies in a range between 0.0 and 3.0 which encompasses $q(t)=5/3=1.67$ expected from the Kolmogorov turbulence in the ``inertial range''\citep{Bieber94, Bieber04}.\par

\begin{figure}[ht!]
\plotone{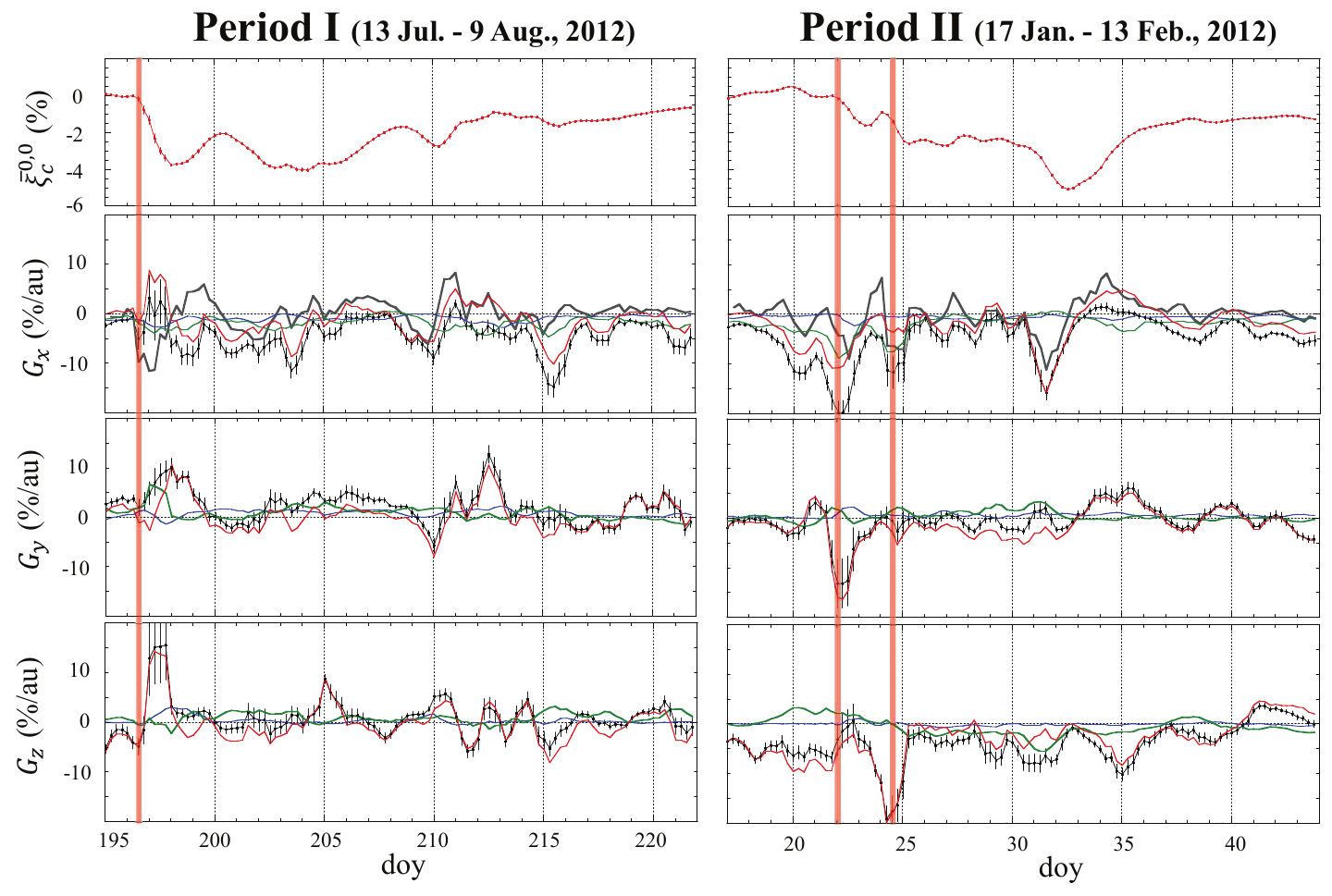}
\caption{Temporal variations of three GSE components of the density gradient vector in Period I (left) and Period II (right) at 15 GV. Top panels show the 24-hour CMAs of the GCR density ($\bar{\xi}_c^{0,0}(t)$) at 15 GV for the reference, while black curves with error bar in the following three panels show the 24-hour CMAs of the GSE x-, y- and z-components of the spatial density gradient vector $\bm G(t)$ derived from the best-fit diurnal anisotropy $\bar{\bm{\xi}}(t)$, respectively. All data are plotted with 6 hour cadence. In each of the bottom three panels, blue, green, red curves display contributions from the parallel diffusion, perpendicular diffusion and diamagnetic drift, respectively, which are calculated by the first, second and third terms of Eq.\ref{G} in the text, respectively. In the second panels, $G_x(t)$ calculated from the time derivative of $\bar{\xi}_c^{0,0}(t)$ is also shown by gray curves (see text). The times of the IP-shock arrival identified by the SSC onset are indicated by vertical orange lines.\label{fig:grad}}
\end{figure}

We finally discuss the spatial gradient of the GCR density. As shown by \citet{Kihara21}, the local spatial density gradient vector ($\bm G(t)$) is given by the local anisotropy vector ($\bm \xi(t)$) as
\begin{equation}
\label{G}
\bm G(t)=\frac{1}{R_L(t)\alpha_{\parallel}(t)}\xi_\parallel(t){\bm b}(t)+\frac{\alpha_{\perp}}{R_L(t)(1+\alpha_{\perp}(t)^2)}\bm{\xi}_\perp(t)+\frac{1}{R_L(t)(1+\alpha_{\perp}(t)^2)}{\bm b}(t)\times\bm{\xi}_\perp(t).
\end{equation}
where ${\bm b}(t)$ is the unit vector along the IMF and $\alpha_{\parallel}$ and $\alpha_{\perp}$ are defined with the local mean-free-paths $\lambda_{\parallel}(t)$ and $\lambda_{\perp}(t)$, as
\begin{equation}
\label{alfa}
\alpha_{\parallel}(t)=\lambda_{\parallel}(t)/R_L(t), \quad \alpha_{\perp}(t)=\lambda_{\perp}(t)/R_L(t).
\end{equation}
Based on $\Lambda_\parallel(t)/R_L(t)$ in Figure \ref{fig:bfparam}(g) exceeding 1.0 and indicating the ``weak-scattering regime'' in most of each period, we simply assume in this paper $\alpha_{\parallel}(t)$ and $\alpha_{\perp}(t)$ to be constant at 7.2 and 0.36, respectively \citep{Wibberenz98, Miyake17, Kihara21}. Since $\bar{\xi}_{\parallel}(t)$ and $\bar{\xi}_{\perp}(t)$ have similar magnitudes in Periods I and II as seen in Figure \ref{fig:bfparam}(d), these constant $\alpha_{\parallel}(t)$ and $\alpha_{\perp}(t)$ result in $\bm G(t)$ dominated by the diamagnetic drift term expressed by the third term on the right hand side of Eq.\ref{G}. Black curves with error bars in the bottom three panels of Figure \ref{fig:grad} show three GSE components of $\bm G(t)$ calculated by this equation, while blue, green and red curves display contributions of the parallel- and perpendicular-diffusions and the diamagnetic drift, respectively. Among these three components, $G_x(t)$ can be also calculated from the time derivative of $\bar{\xi}_c^{0,0}(t)$ as $(1/\bar{V}_{sw}(t))d\bar{\xi}_c^{0,0}(t)/dt$. This $G_x(t)$ is expected to be observed at Earth in the case of the stationary spatial distribution of $\bar{\xi}_c^{0,0}(t)$ passing Earth with $\bar{V}_{sw}(t)$. Gray curves in the second panels indicate $G_x(t)$ calculated in this way. Note that the black curves are deduced from the observed anisotropy ($\bm{\xi}(t)$) by assuming constant $\alpha_{\parallel}$ and $\alpha_{\perp}$ independently from the gray curves calculated from the observed cosmic-ray density and the solar wind velocity. Nevertheless, the overall variation of the black curve in each period looks similar to the variation of the gray curve, although there are some differences seen in magnitudes of local increases and/or decreases. This supports the validation of the $\bm G(t)$ derivation based on the ``weak-scattering regime''. The average level of the black curve ($G_x(t$) is $\sim$5 \%/au lower than that of the gray curve. This is due to the large-scale radial density gradient at Earth which increases near the solar activity maximum period but does not appear in $d\bar{\xi}_c^{0,0}(t)/dt$ at Earth \citep{Munakata14}. In period I, the average $G_x(t)$ is -4.32 \%/au and -0.67 \%/au at 15 GV and 65 GV, respectively, while it is -4.96 \%/au and -0.78 \%/au in Period II. These values are used for $\langle G_r \rangle (=-\langle G_x(t) \rangle)$ in Eq.\ref{Dxi}. It is interesting to note that the average of $G_z(t)$ is negative in Period II, indicating the lower density region in the north of Earth in this period. This is consistent with Enlil movies showing that two later CMEs occurred on DOY 23 and DOY 27 in Period II propagated above the solar equatorial plane, possibly reducing the GCR density in the northern hemisphere.\par

In conclusion, we analyzed two extended GCR decreases in July-August (Period I) and January-February (Period II) in 2012 by using the global analysis method which is improved to suppress the influence of local effects in observations with NMs and MDs each forming a world-wide network. We find that the second step decrease known as the ``phantom FD'' in Period II is rigidity dependent and seen only at lower rigidities. The power-law index of the rigidity spectrum of GCR density ($\bar{\gamma}_0(t)$) varying between -1.0 and -0.5 in Period II caused the ``phantom FD'' at lower rigidities.  In Period I, $\bar{\gamma}_0(t)$ is nearly constant at -1.0 significantly smaller than the index of GCR anisotropy ($\bar{\gamma}_1(t)$) which varies between -0.5 and 0.0 indicating the anisotropy spectrum is systematically harder than that of the density spectrum. Based on the quasi-linear-theory of the GCR diffusion, the observed temporal variations of $\bar{\gamma}_0(t)$ and $\bar{\gamma}_1(t)$ are interpreted in terms of the variation of the power spectrum of the magnetic fluctuation. It is found that the power-law index ($q(t)$) of the magnetic fluctuation spectrum varies between 0.0 and 3.0. In Period I and II, the anisotropy enhancement is seen in both components parallel and perpendicular to the IMF, while the parallel component is consistent with the cosmic-ray streaming along the IMF from the outer heliosphere beyond Earth. Based on a simple model of GCR diffusion along the IMF which connects Earth with the GCR barrier moving radially outward with the average solar wind velocity, we derived the average parallel mean-free-path of the diffusion along the IMF between Earth and the barrier. The eastern angular extension of the barrier necessary for the long duration decreases is roughly estimated to be larger than about 190$^\circ$ at 4.5 au from Sun.\par

\begin{acknowledgments}
This work was supported by JSPS KAKENHI Grant Number JP24k07068. This study is a part of the Science Program of Japanese Antarctic Research Expedition (JARE) Prioritized Research Project (Space environmental changes and atmospheric response explored from the polar cap) which was supported by National Institute of Polar Research (NIPR) under MEXT. It is supported in part by the joint research programs of the National Institute of Polar Research (AJ1007), the Institute for Space-Earth Environmental Research (ISEE), Nagoya University, and the Institute for Cosmic Ray Research (ICRR), University of Tokyo in Japan. This work is also partially supported by a ``Strategic Research Projects'' grant from ROIS (Research Organization of Information and Systems). The observations are supported by Nagoya University with the Nagoya muon detector, by INPE and UFSM with the S\~{a}o Martinho da Serra muon detector, by the Australian Antarctic Division with the Hobart muon detector, and by project SP01/09 of the Research Administration of Kuwait University with the Kuwait City muon detector. N. J. S. thanks the Brazilian Agency - CNPq for the fellowship under grant number 300886/2016-0. EE would like to thank Brazilian funding agencies for research grants FAPESP (2018/21657-1) and CNPq (PQ-301883/2019-0). A. D. L. would like to thank CNPq (PQ 312377/2022-3). We acknowledge the NMDB database (\url{http://www.nmdb.eu}), founded under the European Union's FP7 programme (contract no. 213007) for providing data. We also gratefully acknowledge the NOAA Air Resources Laboratory (ARL) for the provision of GDAS data, which are available at READY website (\url{http://www.ready. noaa.gov}) and used in this paper. The OMNIWeb dataset of the solar wind and IMF parameters is provided by the Goddard Space Flight Center, NASA, USA. The data sets used in this study include those from STEREO-A/PLASTIC and IMPACT and Wind MFI and SWE as retrieved from NASA CDAWeb servers.

\end{acknowledgments}

\newpage

\appendix
\section{Suppression of ``local effects'' in the global analysis}\label{sec:appendix}
In the global analysis of data recorded by many cosmic-ray detectors, special attention needs to be paid to ``local effects'' which are generally different in different detectors depending on the detector's location and/or environment and difficult to correct accurately. In this section, we describe an analysis method which is developed to suppress the influence of such effects.\par
\begin{figure}[ht!]
\plotone{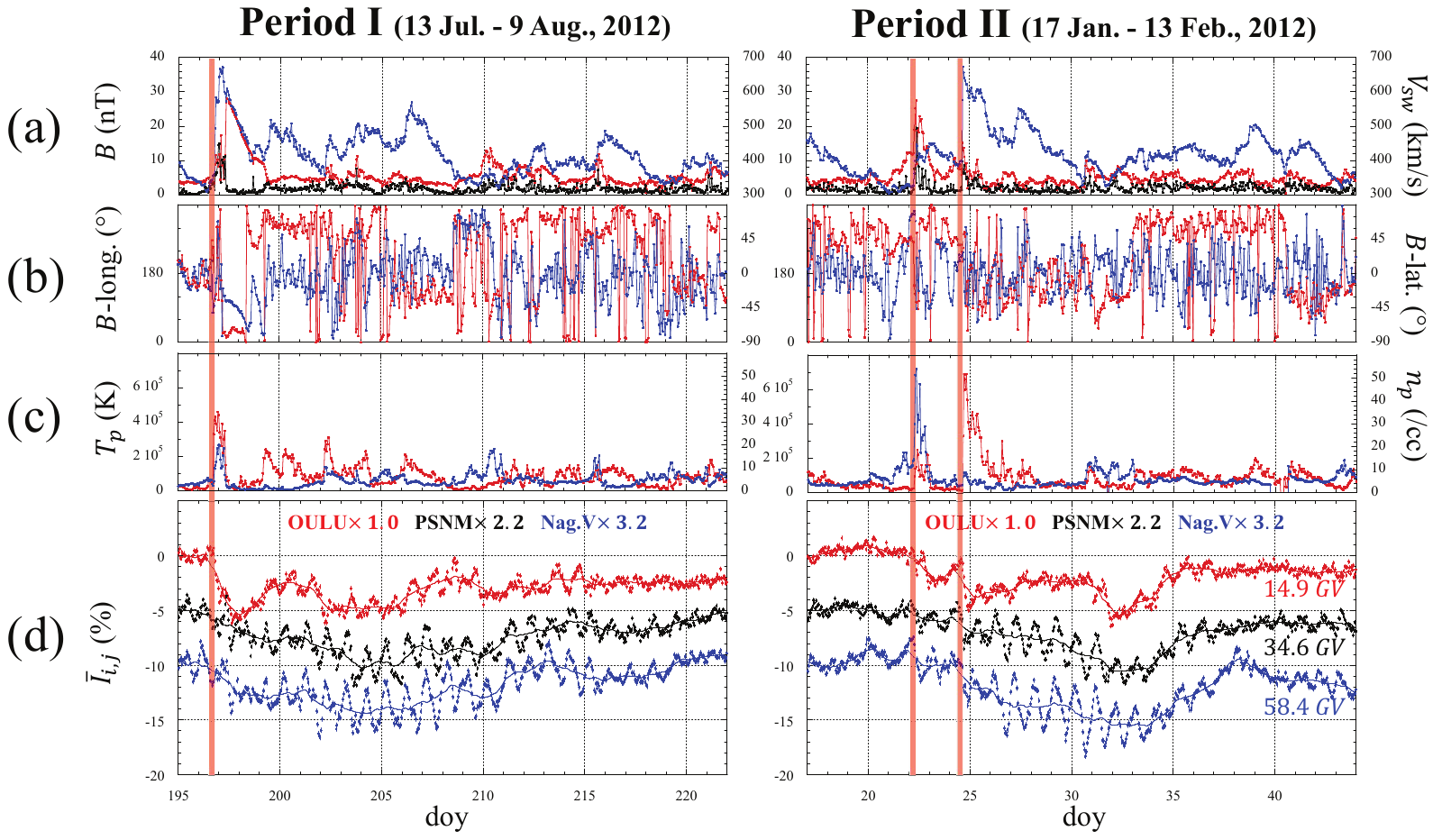}
\caption{Hourly solar wind parameters and cosmic-ray count rates in Period I (left) and Period II (right), each as a function of the day of year (DOY). Panels (a)-(c) show solar wind data; (a) IMF magnitude (red curves) and its fluctuation (black curves) on the left vertical axis and solar wind velocity (blue curves) on the right vertical axis, (b) GSE-longitude (red curves) and latitude (blue curves) of IMF orientation, on the left and right vertical axes, respectively, (c) Proton temperature (red curves) and density (blue curves) on the left and right vertical axes, respectively. Panels (d) show hourly fractional count rates of a sample of two NMs, OULU (red curve) and PSNM (black curve) and the vertical directional channel of Nagoya-MD (blue curve). In panel (d), the count rates of PSNM and Nagoya-MD monitoring higher rigidity GCRs are multiplied by 2.2 and 3.2, respectively and shifted downward for 5 \% and 10 \%, respectively to avoid overlapping. The median primary rigidity ($P_m$) of each detector is indicated below each curve in the right bottom panel. The times of the IP-shock arrival identified by the SSC onset are indicated by vertical orange lines.\label{fig:1hour}}
\end{figure}

We first convert each hourly count rate to the \%-deviation from the average over one solar rotation period and then subtract 24-hour average in the first day of each period which is defined as the starting date of each extended cosmic-ray decrease and indicated by the leftmost orange vertical lines in Figure \ref{fig:1hour}. In other word, we normalize the 24-hour average to 0 \% at this starting date. This is needed, because the \%-deviation from an average over analysis period can be unequally biased depending on the number of data gaps included in the period which is different in different detectors.\par
\begin{figure}[ht!]
\plotone{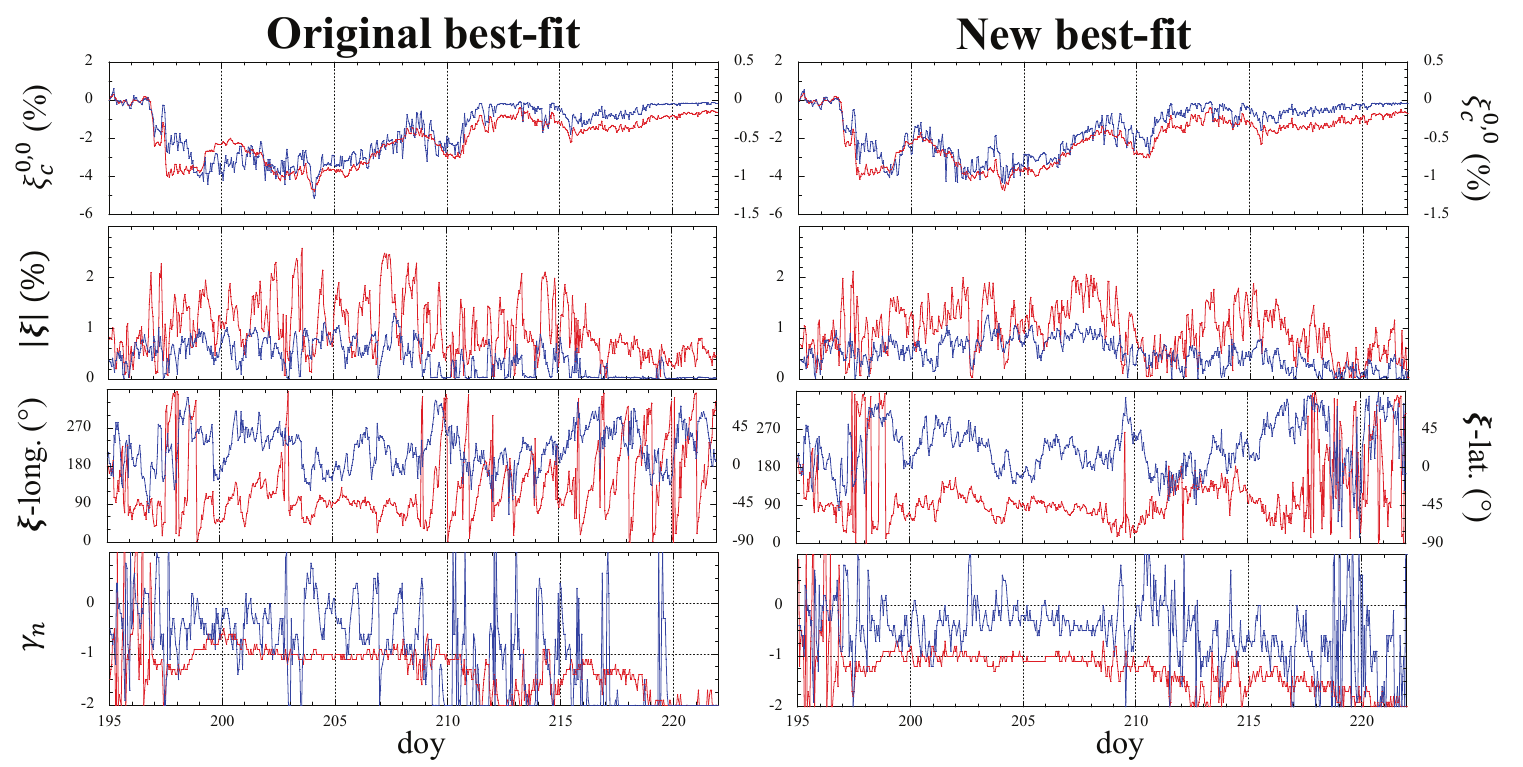}
\caption{Hourly best-fit parameters obtained from the original and new analysis methods for Period I. Left and right panels display parameters obtained from best-fitting to $I^{fit}_{i,j}(t)$ and $I^{nor}_{i,j}(t)$, respectively (see text). Top panels show the best-fit density $\xi_c^{0,0}(t)$ at 15 GV (red curves) and 65 GV (blue curves) on the left and right vertical axes, respectively. Red and blue curves in the second panels show the magnitudes of the anisotropy vector at 15 GV and 65 GV, respectively, while red and blue curves in the third panels show the GSE longitude and latitude of the anisotropy orientation on the left and right vertical axes, respectively. Red and blue curves in the bottom panels show the best-fit $\gamma_0(t)$ and $\gamma_1(t)$, respectively.
\label{fig:old-new}}
\end{figure}
\begin{figure}[ht!]
\plotone{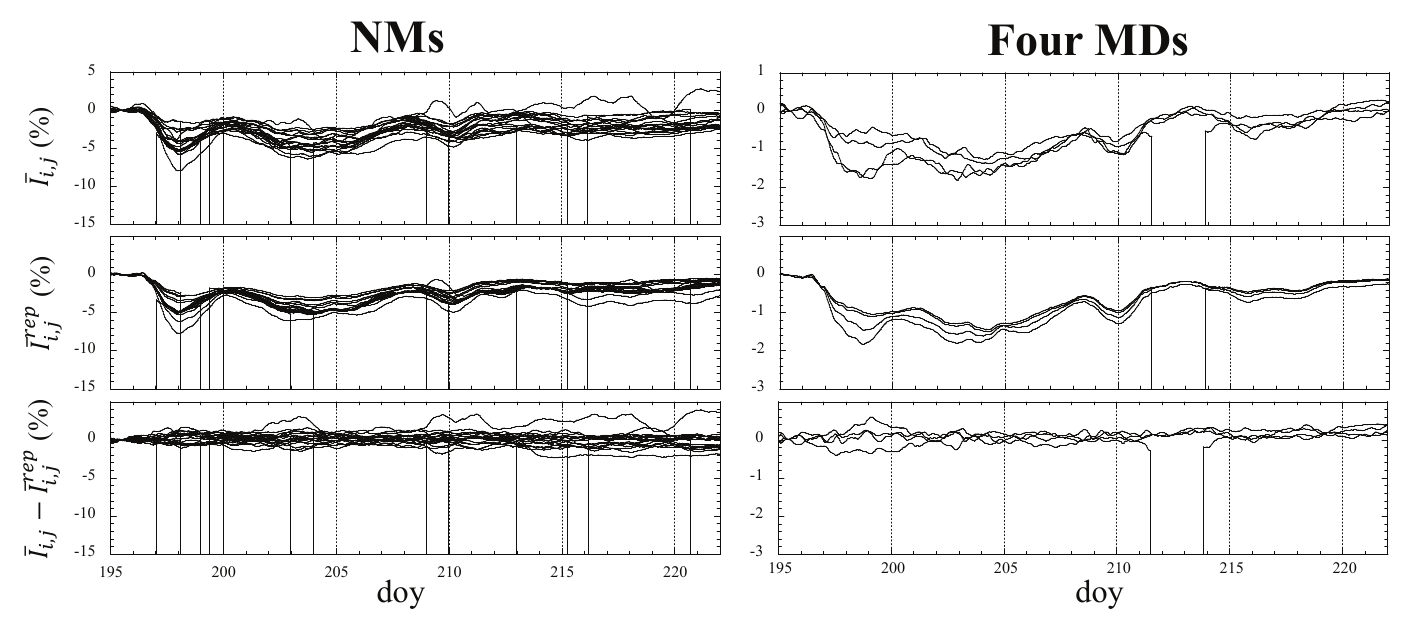}
\caption{24-hour CMAs of GCR count rate observed by NMs (left) and MDs (right) in Period I. Top panels show 24-hour CMAs $\bar{I}_{i,j}(t)$ observed by 26 NMs (left) and four vertical channels of the GMDN (right), while the second panels show $\bar{I}^{rep}_{i,j}$ reproduced from the best-fit ``zonal components'' (see text). The bottom panels show the difference $\bar{I}_{i,j}(t)-\bar{I}^{rep}_{i,j}$ indicating local effect which are generally different in different detectors depending on the detector's location and/or environment.\label{fig:norm}}
\end{figure}

The hourly count rate $I_{i,j}(t)$ in \% calculated in this way is then modeled, as
\begin{equation}
\label{Ifit0}
I^{fit}_{i,j}(t)=\sum_{n=0}^{2}\sum_{m=0}^{n} \{ \xi_c^{n,m}(t) \left( c_{i,j}^{n,m} \cos m\omega t_i - s_{i,j}^{n,m}\sin m\omega t_i \right)+\xi_s^{n,m}(t) \left( s_{i,j}^{n,m} \cos m\omega t_i + c_{i,j}^{n,m}\sin m\omega t_i \right) \}.
\end{equation}
We first use this $I^{fit}_{i,j}(t)$ for the best-fit analysis described in Paper 1. As shown in Paper 1, this best $I^{fit}_{i,j}(t)$ succeeds in reproducing $I_{i,j}(t)$ well. From top to bottom, left panels in Figure \ref{fig:old-new} shows hourly values of, (a) the best-fit density $\xi_c^{0,0}(t_i)$, the amplitude of the first-order anisotropy ($A_1=\sqrt { \{ {\xi_c^{1,0}(t)}^2+{\xi_c^{1,1}(t)}^2+{\xi_s^{1,1}(t)}^2 \} }$), (b) the GSE-longitude and latitude of the anisotropy orientation and (c) $\gamma_0(t)$ and $\gamma_1(t)$, each derived as a function of time during the Period I in a manner described in Paper 1. It is clearly seen that the best-fit anisotropy shows significant 24-hour variations in its amplitude and orientation. Such variation is obviously a flaw in the data and/or analysis method, because there is no reason for the anisotropy in space to vary in 24-hours, the rotation period of Earth. It is found that this spurious variation is arising from the day-to-day variation in the data which is not modeled in $I^{fit}_{i,j}(t)$. If a certain detector records high (low) count for some reason, the best-fit analysis results in the anisotropy from (opposite to) the detector's viewing direction which rotates in space on Earth in 24-hours. This spurious anisotropy, when it is superposed on the real anisotropy, results in 24-hour variations of the anisotropy amplitude and orientation as seen in Figure \ref{fig:old-new}. Top panels of Figure \ref{fig:norm} display the 24-hour central moving averages (CMAs) of $I_{i,j}(t)$ observed by NMs (left) and four vertical channels of MDs (right). Throughout this paper, we calculate the 24-hour CMA of a quantity $X(t)$ as
\begin{equation}
\label{24hCMA}
\bar{X}(t)=\sum_{t-11}^{t+12}{X(t)}/24.
\end{equation}
To see how the observed $\bar{I}_{i,j}(t)$ is reproduced by $I^{fit}_{i,j}(t)$, we plot in middle panels the 24-hour CMAs of the ``zonal components'' calculated as
\begin{equation}
\label{Irep}
\bar{I}^{rep}_{i,j}(t)=\sum_{t-11}^{t+12}{\sum_{n=0}^2\{\xi_c^{n,0}(t)c_{i,j}^{n,0}+\xi_s^{n,0}(t)s_{i,j}^{n,0}\}}/24.
\end{equation}
As seen in the difference ($\bar{I}_{i,j}(t)-\bar{I}^{rep}_{i,j}(t)$) in the bottom panels, $\bar{I}_{i,j}(t)$ in some detector sometimes shows significant deviation from $\bar{I}^{rep}_{i,j}(t)$ in both NM and MD data which cannot be modeled by $I^{fit}_{i,j}(t)$. This deviation is likely arising from the ``local effects'' including the instrumental problems and the environmental effects, such as the snow accumulation effects for NM data 
\citep{Kataoka22} 
and the imperfect correction of the atmospheric temperature effect for MD data \citep{Kato21}. Suppressing the influence of those ``local effects'' is needed particularly in the analysis of the extended CR decrease over a long period. Aiming at such suppression, we use for the best-fit analysis $I^{nor}_{i,j}(t)$ defined as
\begin{equation}
\label{Inorm}
I^{nor}_{i,j}(t)=I_{i,j}(t)(\bar{I}^{rep}_{i,j}(t)/\bar{I}_{i,j}(t)),
\end{equation}
instead of $I^{fit}_{i,j}(t)$. The 24-hour CMA of $I^{nor}_{i,j}(t)$ is normalized to $\bar{I}^{rep}_{i,j}(t)$ calculated from the original best-fit analysis. The right panels of Figure \ref{fig:old-new} display parameters obtained from best-fitting $I^{fit}_{i,j}(t)$ to $I^{nor}_{i,j}(t)$. It is seen that the spurious 24-hour variations are successfully removed in the best-fit anisotropy. In the main text of this paper, we report results obtained from the best-fit analysis using $I^{nor}_{i,j}(t)$. Since this correction method based on the 24-hour CMA may distort shorter period variations of CR density and anisotropy, we would need additional careful analyses when we analyze such shorter variations within 24 hours.

\newpage

\begin{deluxetable}{ccccccccccc}
\tabletypesize{\scriptsize}
\tablewidth{0pt}
\tablenum{1}
\tablecaption{Characteristics of additional six neutron monitors used in this paper\label{tab:CR data}}
\tablehead{
\colhead{name} & \colhead{$\lambda_D (^{\circ})$} & \colhead{$\phi_D (^{\circ})$} & \colhead{alt. (m)} & \colhead{ch-no.} & \colhead{$P_c$ (GV)} & \colhead{cph/$10^4$} & \colhead{$\sigma$ (0.01\%)} & \colhead{$P_m$ (GV)} & \colhead{$\lambda_{asymp} (^{\circ})$} & \colhead{$\phi_{asymp} (^{\circ})$}
}
\startdata
&&&&&&6 NMs\\
\tableline
AATB  &  43.1N &  76.6E  &  3340 & 1 & 6.7 & 540 & 4.3  &  17.1  & 6.9S  & 139.0E\\
JUNG &    46.6N & 7.98E  & 3570 & 1 & 4.5 & 54.0 & 13.6 & 13.5 & 10.0N & 70.0E\\
KIEL   &    54.3N & 10.1E  &   54  & 1  & 2.4 &  64.8  &  12.4 & 15.2 & 11.5N & 54.8E\\
MOSC &    55.5N &  37.3E  & 200  & 1 & 2.4 &  85.5  & 10.8 & 15.1 & 14.3N & 77.8E\\
NEWK  &    39.7N & 284.3E &    50 & 1 &  2.4 & 34.1 &  17.1 &  15.3 & 1.84S & 338.4E\\
ROME  &  41.9N  &  12.5E  &      0 & 1 & 6.3 &  54.9 & 13.5 &  19.5 &  0.9N  &  71.5E\\
\tableline
\enddata
\tablecomments{Each column lists the detector name, geographic longitude ($\phi_D$) and latitude ($\lambda_D$), altitude of detector's location, number of directional channels available from the detector, geomagnetic cut-off rigidity ($P_c$) for each directional channel, average hourly count rate, count rate error ($\sigma$), median rigidity of primary GCRs ($P_m$), geographic longitude ($\phi_{\rm asymp}$) and latitude ($\lambda_{\rm asymp}$) of the asymptotic viewing direction outside the magnetosphere. $P_m$ is calculated by using the response function of each detector to primary GCRs, while $\lambda_{\rm asymp}$ and $\phi_{\rm asymp}$ are calculated by tracing orbits of GCRs with $P_m$ in the model magnetosphere (see Paper 1).}
\end{deluxetable}

\newpage

\newpage

\newpage

\newpage

\newpage

\newpage

\newpage

\end{document}